# Effect of uniaxial compressive stress on polarization switching and domain wall formation in tetragonal phase BaTiO$_3$ via machine learning potential

*Po-Yen Chen[1\*], Teruyasu Mizoguchi[1,2\*]*.

AUTHOR ADDRESS

[1]Department of Materials Engineering, the University of Tokyo, Tokyo, Japan

[2]Institute of Industrial Science, the University of Tokyo, Tokyo, Japan.

AUTHOR INFORMATION

**Corresponding Author**

poyen@iis.u-tokyo.ac.jp, teru@iis.u-tokyo.ac.jp

ABSTRACT

Ferroelectric materials such as $BaTiO_3$ exhibit spontaneous polarization that can be reoriented by an external electric field, forming the basis of various memory, actuator, and sensor applications. The polarization switching behavior, however, is strongly influenced by mechanical boundary conditions due to the intrinsic electromechanical coupling in ferroelectrics. In this study, we employ a machine learning interatomic potential to investigate the effect of uniaxial compressive stress on polarization switching and domain wall evolution in the tetragonal phase of $BaTiO_3$. This study revealed a critical stress about 120 MPa which 90° polarization switching occurs. Beyond the critical stress, larger supercells exhibit lower activation energies for polarization switching with 180-degree domain wall formation and weaker constraints from periodic boundary conditions, thereby facilitating domain-wall formation. Besides, Increasing compressive stress reduces both the remnant polarization and the coercive field, while a double hysteresis loop emerges at a stress level of 80 MPa. These findings provide atomistic insights into stress-controlled ferroelectric switching and highlight the crucial role of mechanical loading in designing reliable ferroelectric devices.

# 1. Introduction

Barium titanate (BaTiO$_3$), a prototypical ferroelectric perovskite, exhibits strong dielectric and piezoelectric properties that have enabled wide applications in capacitors, sensors, and non-volatile memories[1-3]. In particular, the polarization switching behavior, which involves the reorientation of electric dipoles under an external electric field or mechanical stress[4], plays a central role in determining ferroelectric performance. Recent studies have shown that mechanical stress can strongly influence the polarization switching process by altering the local atomic structure, energy landscape, and domain configuration[5, 6]. Depending on the stress magnitude and direction, the coercive field, switching pathway, and kinetics of domain evolution can be significantly modified[7-9]. Moreover, stress can affect the electric field responses of BaTiO$_3$-based materials, such as the shape and area of the polarization–electric field (P–E) hysteresis loop, reflecting the coupling between mechanical stress and electric field[9, 10]. Also, large stresses may promote or hinder polarization reversal and even induce the formation of 180-degree domain walls (DWs)[8]. While isostatic pressure applies uniform stress in all directions, preserving the overall lattice symmetry, uniaxial stress acts along a single direction and can induce anisotropic lattice deformation and direction-specific responses, such as dielectric properties and polarization switching[11, 12].

Therefore, bridging the macroscopic effects of mechanical stress with atomic-scale behavior, we utilized machine-learning potentials to perform molecular dynamics simulations, allowing us to study polarization switching dynamics, the evolution of ferroelectric domains, and stress-induced structural changes in tetragonal BaTiO$_3$. This approach enables a detailed understanding of how uniaxial stress governs the atomic-level mechanisms underlying polarization switching and ferroelectric performance.

To achieve detailed insights on the stress effects on the domain wall dynamics, However, conventional approaches such as the phase-field method and the Landau-Ginzburg-Devonshire (LGD) thermodynamic potential are mainly employed to study the coupling between stress and polarization because of their high efficiency and suitability for modeling μm-scale systems[6, 13, 14]. Nevertheless, these methods cannot capture atomic-scale phenomena and are often limited to specific material parameters, which restricts their applicability for understanding the microscopic mechanisms under various conditions.

Although the core–shell model for the atomic scale simulation has been used in several recent studies, its limited accuracy may affect the quantitative reliability of their results[15]. To overcome these limitations, we employed machine learning potential-based molecular dynamics (MLP-MD) simulations[16-18], which enable atomic-scale, highly accurate, and efficient modeling of ferroelectric behavior. Unlike the phase-field and LGD approaches, MLP-MD can explicitly describe the influence of stress and electric fields on atomic displacements and local polarization. Furthermore, since MLPs exhibit accuracy comparable to ab initio molecular dynamics (AIMD) and outperform the core-shell model in predicting energies and forces, this approach can faithfully reproduce atomic trajectories and provide deeper insights into the dynamic processes governing polarization switching in $BaTiO_3$.

In this study, we systematically investigate the influence of mechanical stress on the ferroelectric behavior of $BaTiO_3$ through three aspects. First, we examine the uniaxial stress-induced polarization switching process to clarify how mechanical loading affects the atomic-scale switching dynamics. Next, we analyze the impact of uniaxial compressive stress on the P–E hysteresis loop to reveal its effect on the hysteresis behavior. Finally, we explore the formation mechanism of stress-induced DWs to understand how stress drives structural and polarization reconfiguration in BaTiO$_3$.

## 2. Method

### 2.1. MLP-MD simulation setting

In this study, MD simulations were performed using the NPT ensemble implemented in the Atomic Simulation Environment (ASE) package with a time step of 1 fs. The system temperature was controlled by a Nosé–Hoover thermostat with a time constant of 25 fs, while the pressure was maintained at 0 GPa using a Parrinello–Rahman barostat with a time constant of 75 fs. The MLP model employed in this work was developed in our previous study, where the training database was generated using DFT calculations with the PBEsol functional[19]. Specifically, the model was obtained by fine-tuning the MACE-MP-0 model[20] on our BaTiO$_3$ database consisting of 4,045 structural configurations. As a result, the fine-tuned MLP can accurately and efficiently describe the atomic interactions and ferroelectric behavior of BaTiO$_3$ systems. In our previous validation, the model demonstrated excellent predictive accuracy for both energies and atomic forces. Moreover, it successfully reproduced the temperature-dependent phase transition sequence with transition temperatures in close agreement with experimental measurements. In addition, the model accurately captured the local atomic environments of liquid and amorphous phases, indicating its robustness in describing highly distorted configurations.

Furthermore, to evaluate the capability of our MLP to reproduce stress-induced responses, we calculated the elastic constants using slightly strained structures with −0.5% and +0.5% deformation. The elastic constants were extracted by fitting the strain–energy relationship to a second-order polynomial. As summarized in Table 1, the results were benchmarked against previous DFT calculations[21-24] and available experimental measurements[23, 25-29]. Overall, the elastic constants predicted by our MLP are in good agreement with those obtained from simulation results[21-24], especially with PBEsol results, although the values of $C_{11}$ and $C_{44}$ are slightly higher than the experimental data[23, 25-29]. Notably, the predicted elastic constants closely match those reported in studies employing the PBEsol functional[21, 30], indicating that the proposed MLP reliably captures DFT-level mechanical properties and accurately describes the response of BaTiO$_3$ under small strains.

Table 1. Comparison of the elastic constants obtained in this work with previous simulation results[21-24] and experimental values[23, 25-29].

|  | $C_{11}$ | $C_{12}$ | $C_{13}$ | $C_{33}$ | $C_{44}$ | $C_{66}$ |
| --- | --- | --- | --- | --- | --- | --- |
| This study | 321 | 125 | 113 | 164 | 115 | 125 |
| GGA-PBEsol[21] | 281 | 98 | 89 | 128 | 111 | 121 |
| GGA-PBE[22] | 246 | 101 | 104 | 159 | 68 | 119 |
| GGA-PW[23] | 255 | 101 | 104 | 159 | 68 | 119 |
| LDA[24] | 300 | 109 | 90 | 149 | 124 | 128 |
| Experiment[23, 25-29] | 211~275 | 107~179 | 104~151 | 126~165 | 42~64 | 93~134 |

**2.2 BEC calculation and BEC prediction model**

To calculate the P–E hysteresis loop of BaTiO$_3$, the Born effective charges (BECs) of BaTiO$_3$ are required[31, 32]. The reference BECs were obtained using density-functional perturbation theory (DFPT)[33, 34] as implemented in VASP[35-37]. The structures included in the database described in our previous study were calculated using the PBEsol exchange–correlation functional with a plane-wave cutoff energy of 500 eV and a Γ-centered 5×5×5 k-point mesh. Static calculations were performed with an electronic convergence criterion of 1×10⁻⁶ eV.

To accelerate the BEC evaluation during the hysteresis loop simulations, we employed the Equivar_eval model developed by Kutana's work[38, 39]. This model can accurately and efficiently predict BEC values directly from structural information without additional first-principles calculations. Although the computational time and cost of the GNN model increase with the number of atoms, it still significantly enhances the efficiency for large-scale systems[40, 41]. To further enhance its accuracy for BaTiO$_3$ systems, we fine-tuned the pretrained Equivar_eval model, BM1 model, using our own BEC database described above, with a training–validation–test split ratio of 6:2:2. The comparison between the pretrained and fine-tuned models is shown in Figure 1, where the mean absolute error (MAE) was reduced from 0.1691 to 0.0697, demonstrating a significant improvement in BEC prediction accuracy for BaTiO$_3$. Therefore, the fine-tuned Equivar_eval model was employed in this study to simulate polarization change and to perform the subsequent MLP-MD simulations.

## 2.3 Polarization and Ti-displacement calculation

The polarization was computed using the expression of $P = \frac{e}{\Omega}\sum_{\kappa} Z_{\kappa}^* \cdot \vec{u_{\kappa}}$, where $e$ is the elementary charge, $\Omega$ is the volume of the structure, $Z_{\kappa}^*$ is the BEC tensor for the atom κ obtained by the finetuned Equivar_eval model, and $\vec{u_{\kappa}}$ is the displacement of atom $\kappa$ relative to the reference cubic-phase BaTiO$_3$ structure.

In addition, the Ti displacement was defined as the vector from the Ti atom to the center of mass of its six surrounding oxygen atoms, following the method described in our previous study on BaTiO$_3$ MLP. The Ti displacement provides an intuitive measure of local polarization changes within the structure and can be used to analyze polarization switching at the atomic level.

**2.4 P-E hysteresis loop simulation**

In this study, we employed MLP-MD simulations to investigate the P–E hysteresis loop of BaTiO$_3$. MD simulations were conducted on an 8×8×8 supercell of tetragonal-phase BaTiO$_3$. The initial configuration was prepared by equilibrating the system via a 20 ps MD simulation at 250 K. Since the MLP does not inherently account for the effect of an external electric field, we developed a MD framework to incorporate this effect explicitly. After obtaining the BECs from the fine-tuned Equivar_eval model, the force due to the external electric field on atom κ can be calculated as $F_{ext}^{(\kappa)} = |e|\mathcal{E}_\beta Z^*_{\kappa,\beta\alpha}$, where $\mathcal{E}$ is the applied electric field, $Z^*$ is the BEC tensor. The total force on each atom is then obtained as $F_{total} = F_{MLP} + F_{ext}$, where $F_{MLP}$ is the force calculated by MLP. Using this framework, we applied a triangular-wave electric field with a maximum amplitude of $\mathcal{E}_{max} = 100$ kV/cm and a frequency of 2.5 GHz for the P-E hysteresis loop simulations, while maintaining the system temperature at 250 K. The P–E hysteresis loop was then constructed from the time evolution of the total polarization. MD simulations were performed with a time step of 1 fs for a total duration sufficient to capture at least several full cycles of the applied electric field.

**3. Results**

## 3.1 Stress-induced polarization switching

To investigate the effect of mechanical loading on atomic-scale polarization switching dynamics, we first conducted a stress-dependent lattice constant analysis to examine structural changes under various constant stresses. In our previous study, we found that the 8×8×8 BaTiO$_3$ supercell containing 2,560 atoms, provides a converged phase transition temperature, with the tetragonal phase remaining stable between 225 and 290 K[19]. Therefore, in this analysis, uniaxial compressive stress was applied along the polarization direction (c-axis) of an 8×8×8 supercell of tetragonal BaTiO$_3$ as shown in Figure 2a, with stress magnitudes of 0, 40, 80, 120, 160, 320, 480, 640, and 800 MPa. All simulations were performed at 250 K for 20 ps. Figure 2b shows the variation of the lattice constants along the a- (green), b- (orange), and c- (blue) axes under uniaxial compressive stress. It can be observed that the lattice constants initially exhibit a linear dependence on the applied stress in the range of 0 to 80 MPa. From this linear region of the stress–strain response, the Poisson's ratio of tetragonal BaTiO$_3$ can be extracted, yielding a value of 0.32. This result is in similar with experimental measurements[23, 25-27], further validating the reliability of our MLP-based MD simulations in capturing the elastic response of BaTiO$_3$ under small deformations. After a pronounced change around 120 MPa, the lattice constants again display an approximately linear relationship with stress. Moreover, the b-axis lattice constant surpasses that of the c-axis and becomes the largest among the three, while the c-axis lattice constant decreases to a value similar to that of the a-axis. This behavior indicates that polarization switching occurs when the uniaxial compressive stress is applied.

In previous studies for phase field model and LGD approaches have indicated that the 90-degree polarization switching would happen with the mechanical stress is loading over the coercive stress[6, 42]. For the experimental observation, Li et al. also utilize the compressive stress to achieve the 90-degree polarization switching[43]. Therefore, in this case, our observation for the 90° polarization switching has a nice agreement with the simulation from the LGD theory.

Besides the lattice constants and polarization, mechanical stress can also influence several other material properties of $BaTiO_3$, such as the hysteresis behavior and DW formation. Previous studies based on the Wang's phase-field simulations suggested that, in the absence of an external bias voltage, uniaxial compressive stress exceeding the coercive stress can induce random polarization switching, leading to the formation of 180° DWs[44]. Furthermore, LGD theory have revealed pronounced electromechanical coupling effects in ferroelectric $BaTiO_3$[6, 9]. Motivated by these theoretical insights, we further investigated the effects of both low and high uniaxial compressive stresses on $BaTiO_3$ to examine their influence on its ferroelectric behavior and DW formation.

**3.2 Stress-induced DW formation**

Next, we aim to investigate the stress-induced DW formation. Although several previous studies have discussed various DW properties, such as formation energy and DW motion, using first-principles calculations, phase-field models, classical force fields, and machine-learning-based potentials, these works primarily focused on systems that already possess pre-existing DW structures[45-48]. Consequently, it remains challenging to directly explore the DW formation process and its governing factors. Therefore, in this study, we applied uniaxial compressive stress to induce DW formation and to elucidate the key factors controlling this stress-driven process. In the previous sections, we demonstrated that a stress exceeding the coercive stress can induce 90° polarization switching. Moreover, earlier phase-field simulations have suggested that random polarization switching may lead to the formation of 180° DWs[44]. To further examine this phenomenon, we employed a larger simulation cell and applied a high uniaxial compressive stress. Specifically, as illustrated in Figure 3a, a tetragonal $BaTiO_3$ supercell consisting of 8×8×32 unit cells (3.2 nm × 3.2 nm × 12.8 nm, containing 10,240 atoms) was constructed and equilibrated at 250 K for 20 ps. Subsequently, a uniaxial compressive stress of 800 MPa was applied along the c-axis (polarization direction) for another 20 ps. Figure 3b presents the projected Ti displacements on the bc-plane. Initially, the structure exhibits a single-domain state with polarization along the c-axis. Under compressive stress, however, the Ti displacement vectors reorient toward the a/b directions, accompanied by the formation of 180° DWs. This observation is in good agreement with previous phase-field studies and confirms the feasibility of stress-induced domain-wall formation in $BaTiO_3$.

To characterize the nature of the stress-induced DWs, we analyzed the Ti displacements along the three crystallographic axes as a function of position along the c-axis within the DW region. As shown in Figure 4, the Ti displacements along the b-axis exhibit large magnitudes of approximately 0.15 Å, whereas those along the other two axes remain below 0.05 Å, indicating that the polarization is primarily oriented along the b-axis. A distinct reversal of the b-axis Ti displacement is observed, marking the occurrence of a DW. The DW position is defined as the point where the b-axis Ti displacement crosses zero. Based on this criterion, we quantified the DW distance as the minimum separation between adjacent DWs, and the DW width as the region where the b-axis Ti displacement varies from –0.1 Å to 0.1 Å. Furthermore, analysis of the Ti displacement variations along the c-axis shows that the displacements along the a- and c-axes exhibit no significant changes, indicating that the polarization does not rotate within the DW region. Therefore, the DW formed under uniaxial compressive stress can be classified as an Ising-type DW. This observation is consistent with previous theoretical predictions based on the Ginzburg–LGD framework[8, 49, 50], confirming the reliability of the stress-induced DW configuration obtained from our MLP-MD simulations.

To further elucidate the factors governing stress-induced DW formation, we systematically investigated the effects of applied stress and supercell length. To ensure structural stability, the lattice parameters along the a and b axes were fixed at 32 Å, corresponding to an 8×8 supercell, which maintained the tetragonal $BaTiO_3$ phase throughout all simulations. To examine the stress dependence, uniaxial compressive stresses of 80, 160, 320, 480, 640, and 800 MPa were applied along the c-axis to a 128 Å-long supercell at 250 K for 20 ps under the NPT ensemble. Each condition was simulated ten times to obtain statistically reliable data on DW formation. The changes in DW formation probability, average DW distance, and average DW width are summarized in Figure 5. As shown in Figure 5a, when the applied stress exceeds 160 MPa, the DW formation probability markedly increases to approximately 0.7–0.9, indicating that once the stress surpasses a critical value, stress-induced DW formation becomes energetically favorable and stable. This trend is similar to the polarization switching behavior observed in the 8×8×8 supercell, highlighting the critical role of the threshold stress in facilitating DW formation through polarization switching. Regarding the DW distance shown in Figure 5b, it shows a similar value of around 40 Å and the no clear correlation was observed with increasing stress, suggesting that the magnitude of compressive stress does not significantly influence the spacing between DWs. In contrast, as shown in Figure 5c, the DW width decreases with increasing compressive stress and eventually stabilizes at approximately 5.5 Å. This value is comparable to the DW widths obtained from molecular dynamics simulations combined with Monte Carlo methods and from effective Hamiltonian calculations without applied stress[51, 52]. It also shows good agreement with first-principles calculations and LGD model, which suggest the Ising DW width of less than 1 nm[48, 53, [54]. This result indicates that stress has a pronounced influence on the DW width, although an intrinsic limit of the polarization gradient still exists under compressive stress. These findings highlight the crucial role of the critical stress in promoting polarization switching and stabilizing domain-wall formation. Beyond this critical point, further increases in stress have only a minor influence on the characteristics of the domain walls.

Next, we investigated the effect of supercell length on DW formation. Supercells with c-axis lengths of 64, 80, 96, 112, and 128 Å, corresponding to 16, 20, 24, 28, 32 supercells along with c-axis, were constructed and equilibrated at 250 K for 20 ps in the NPT ensemble. A uniaxial compressive stress of 800 MPa was then applied along the c-axis to induce DW formation as similar to the scheme in Figure 6a, and each configuration was simulated ten times to evaluate the probability of DW formation, the average DW distance, and the DW width.

First, regarding the dependence of DW formation probability on the supercell length, as shown in Figure 6a, the probability remains zero when the supercell length is 64 Å, indicating that a minimum spatial extent is required for the domain wall to form. Moreover, when the length reaches 80 Å or above, the DW formation probability increases with increasing supercell length. This trend suggests that larger simulation cells facilitate DW formation, possibly due to the reduced constraint arising from the periodic boundary conditions of the initial polarization distribution. As shown in Figure S1, the polarization remains nearly uniform across the structure. This uniformity originates from the strong constraints imposed by periodic boundary conditions, which suppress polarization fluctuations and effectively restrict the system to a single-domain configuration. In Section 4.1, we will discuss this phenomenon in energy-based analysis. In contrast, larger supercells allow broader and varying fluctuations in Ti displacement, since PBC only constrains the distant boundaries while the interior can accommodate gradual spatial variations.

For the DW distance, Figure 6b shows that the distance increases significantly with increasing supercell length, together with a larger standard deviation. This behavior implies that longer supercells provide sufficient spatial freedom for the spontaneous development of multiple or more widely separated DWs, thereby reducing the artificial constraints imposed by the periodic boundaries. In contrast, as shown in Figure 6c, the DW width, unlike the stress effect, remains nearly constant at approximately 5.5 Å regardless of the supercell length. This result suggests that the DW width is an intrinsic characteristic primarily governed by the applied uniaxial compressive stress, rather than by finite-size effects.

**3.3 Stress effect on the hysteresis loop**

In the following section, the stress effects on the P-E hysteresis behavior of $BaTiO_3$ are investigated. Before proceeding, it is necessary to first examine the P–E hysteresis loop under normal conditions as a reference for the unstressed state. In this case, we focus on the 8×8×8 $BaTiO_3$ supercell since it is the minimum supercell for the stable phase transition temperature. The P–E hysteresis loop of the 8×8×8 tetragonal $BaTiO_3$ supercell was calculated at 250 K using MD simulations in the NPT ensemble to represent the normal, unstressed state. As shown in Figure 7, the polarization along the a-, b-, and c-axes was monitored under a triangular electric field with a maximum amplitude of 100 kV/cm. The results show that our method can successfully reproduce the hysteresis loop for tetragonal $BaTiO_3$, with a remnant polarization of approximately 20 μC/cm² and a coercive field of about 50 kV/cm. The remnant polarization is in good agreement with previous simulations and experimental results[6, 15, 55, 56], whereas the calculated coercive field is only one-third of the typical values obtained from core-shell model[15] and LGD theory[6].

Considering that the BEC calculated by the core-shell model is often underestimated, our results indicate that the present MLP-MD approach provides a more accurate description of polarization behavior than core-shell simulations. However, the coercive field in our results is still roughly one order of magnitude larger than experimental values[55, 57, 58], which may be attributed to internal stress caused by crystalline imperfections in BaTiO$_3$, as discussed in Azuma's study[15]. Furthermore, the dielectric constant $\varepsilon_c$ was evaluated from the linear region of the hysteresis loop (−10 to 10 kV/cm) and was found to be approximately 153 in averaged, in good agreement with experimental data at room temperature[59]. This further confirms the reliability and feasibility of our method for simulating ferroelectric properties of BaTiO$_3$.

Besides, we utilized the linear response between –20 kV/cm and 20 kV/cm along the a-axis, as shown in Figure S2, and obtained a dielectric constant ($\varepsilon_a$) value of 1070. This value is comparable to the $\varepsilon_\perp$ of the tetragonal phase reported in Gigli's study[60]. However, experimental measurements[61] have shown that the $\varepsilon_a$ of BaTiO$_3$ typically lies between 3000 and 4000, indicating a significant discrepancy between Gigli's results and ours. We attribute this difference mainly to the use of the PBEsol exchange–correlation (XC) functional in the DFT calculations. As reported in previous studies, different XC functionals yield different c/a ratios for BaTiO$_3$, with PBEsol generally producing a larger c/a value than the experimental one[62-66]. This overestimation of c/a likely restricts the mobility of Ti atoms, particularly their displacement along the a-axis, thereby reducing the calculated dielectric constant in both Gigli's and our studies. Nevertheless, the obtained dielectric constant is within the same order of magnitude and adequately reproduces the qualitative dielectric behavior of tetragonal BaTiO$_3$.

In addition, as shown in Figure 7, the polarization along the a- and b-axes exhibits sudden pulses of similar magnitude to the original c-axis polarization during polarization switching. This behavior indicates that polarization switching in BaTiO$_3$ occurs via a two-step process with the c → ± a/b → − c. This phenomenon has been observed in several simulation studies, including those based on the core-shell model[15] and LGD thermodynamic potential[6]. Experimental studies have also reported two-step polarization switching under both small and large electric fields[67-69]. These results indicate that our MLP-MD approach can not only reproduce the overall hysteresis loop but also capture the detailed atomic-scale polarization switching behavior of BaTiO$_3$.

Next, to investigate the stress effect on electric field response, uniaxial compressive stresses of 80 MPa (orange) and 160 MPa (green) were applied to the 8×8×8 BaTiO$_3$ supercell, while keeping the same electric field amplitude and frequency. The results are compared with the stress-free case (blue) to evaluate the effects of small and large compressive stresses on the P–E characteristics, as shown in Figure 8. Under a large compressive stress of 160 MPa, the P–E curve exhibits a nearly linear polarization–electric field relationship, indicating a paraelectric-like behavior. In contrast, a small compressive stress of 80 MPa leads to the emergence of a double-hysteresis loop, characterized by a distinct two-step polarization switching. Unlike in antiferroelectric materials, this loop still retains a finite remnant polarization at zero electric field. A similar phenomenon was previously reported by Wu et al. using the LGD approach, suggesting that such double-hysteresis loops can occur near the Curie temperature or under moderate uniaxial compressive stress[9]. They attributed this behavior to the competition between the c-axis and a/b-axis polarization phases, where increasing compressive stress facilitates an electric-field-induced pseudo-phase transition that gives rise to the double-hysteresis feature. The detailed origin of double-hysteresis behavior will be discussed in Section 4.2.

Although we did not perform the hysteresis loop simulation for the stress-induced domain wall structure due to the high computational cost of the GNN-based Equivar_eval model, we expect that the P–E curve of the stress-induced structure, which requires stress levels exceeding 160 MPa, would be similar to that of the paraelectric phase.

## 4. Discussions

### 4.1 Origin of the domain wall formation probability difference by the supercell length effect

To elucidate the determining factors governing the DW formation probability with respect to supercell length, we evaluated the DW formation energy. The DW formation energy was calculated using $E_{f,DW} = E_{DW,ave} - E_{no\_DW,ave}$, where $E_{DW,ave}$, $E_{no\_DW,ave}$ represent the time-averaged potential energies over 2 ps for the structures with and without DWs, respectively, at a given supercell length and 800 MPa. As shown in Figure 9a, the variation in supercell length exerts little influence on the DW formation energy, which remains generally positive value of 160 MPa, which have a good agreement with Monte Carlo simulation using an effective Hamiltonian method[52]. However, this cannot process the information why the longer supercell can have higher probability to form DW structures.

To further examine the origin of the observed length dependence, we evaluated the potential energy evolution during the stress-induced polarization switching process for supercells of different lengths. As shown in Figure S3, when uniaxial compressive stress is applied, the 90° polarization rotation from the c-axis to the ab-plane occurs within approximately 2 ps. After this point, both the polarization reversal and the emergence of DW are observed, indicating that this dynamical pathway simultaneously governs DW nucleation. Therefore, a reduction in the activation energy along this pathway implies an increased likelihood of DW formation. The activation energy for stress-induced polarization switching was defined as the maximum value of the potential energy within the first 2 ps of the switching event, measured relative to the initial state. The length-dependent activation energies are summarized in Figure 9b. The activation barriers for polarization switching without DW formation (orange curve) remain nearly independent of supercell length with the value of around 1.1 meV/Å, suggesting that single-domain switching is not strongly size-dependent. In contrast, the activation energies with DW formation (blue curve) decrease systematically as the supercell length increases, eventually becoming lower than the single-domain switching barrier. This trend explains the enhanced DW formation probability in larger supercells shown in Figure 6a, and the particularly high DW nucleation probability observed for the 128 Å (32-cell) supercell. These findings indicate that the reduced activation energy for DW-mediated switching in larger supercells is a key factor enabling stress-induced DW formation. This length-dependent reduction in the activation barrier provides a mechanistic explanation for the enhanced DW formation probability under mechanical loading.

### 4.2 Origin of the double-hysteresis loop

About the double-hysteresis loop observed in Section 3.3, it is mainly derived from the coercive electric fields of two-step polarization switching. In a previous study, Li et al. demonstrated within the LGD framework that the coercive electric fields associated with the two-step polarization switching are highly sensitive to the applied uniaxial stress[6]. In the absence of stress, the coercive field for the second switching process (from the a/b-polarized to the c-polarized state) is lower than that of the first switching (from the c-polarized to the a/b-polarized state), indicating that the overall coercive field in BaTiO$_3$ is dominated by the first transition, while the second occurs almost instantaneously afterward. Upon applying uniaxial compressive stress, the coercive field of the first switching decreases, whereas that of the second increases. When the compressive stress reaches the critical value (~100 MPa in their study), the coercive fields of these two processes intersect, leading to distinct switching events at different electric fields and resulting in the appearance of a double-hysteresis loop in the P–E curve. In addition, their results showed that under very large compressive stress (~225 MPa in their study), the coercive field for the first switching approaches zero, which can also account for the stress-induced polarization suppression and the paraelectric-like behavior observed in our P–E curves at 160 MPa.

Although no direct experimental observation of mechanically stress-induced double-hysteresis loops in BaTiO$_3$ has been reported, similar behaviors have been observed in doped BaTiO$_3$ systems such as Bi–BCT, K-doped BaTiO$_3$, and aged BaTiO$_3$ crystals[70-72]. In addition, Zhang *et al.* demonstrated through phase-field simulations that dislocations can also induce double-hysteresis loops by generating local stress fields[73]. These findings suggest that intrinsic stresses arising from dopants or defects may lead to the appearance of double-hysteresis loops, thereby supporting the notion that mechanical stress can significantly influence the hysteresis behavior of BaTiO$_3$ systems.

## 5. Conclusion

In conclusion, we have investigated the effects of uniaxial compressive stress on the polarization switching behavior of bulk tetragonal BaTiO$_3$ using MLP-MD. We further examined how stress influences hysteresis behavior and domain-wall formation. A critical stress of approximately 120 MPa was identified, beyond which a 90° polarization switching and corresponding changes in elastic properties occur, in qualitative agreement with trends predicted by LGD theory and observed in experimental studies[6, 42, 43]. Regarding DW formation, DWs appear only when the stress exceeds the critical value, and the DW width decreases with increasing stress. Furthermore, increasing the supercell size lowers the activation energy and weakens the artificial constraints imposed by periodic boundary conditions, thereby facilitating domain-wall formation during stress-induced polarization switching.

Besides, when the applied stress exceeds this critical value, BaTiO$_3$ exhibits paraelectric behavior along the original polarization direction. Below the critical stress, both the remnant polarization and coercive field decrease with increasing stress, and a double hysteresis loop is observed at 80 MPa, similar to LGD theory[6] and experimental observations in defect-containing samples[70-73]. Overall, our findings provide a comprehensive understanding of the effects of uniaxial compressive stress on the polarization behavior and DW evolution in BaTiO$_3$. These insights may offer valuable guidance for the design of piezoelectric and ferroelectric devices and for controlling stress-induced domain structures in perovskite oxides.

SUPPLEMENTARY MATERIAL

The supplementary material contains the c position-dependent Ti-displacement for 8x8x8 and 32x8x8 supercell without the stress, the step-dependent energy change during polarization switching process, and the P-E curve for linear response region for BaTIO$_3$ along a-axis.

AUTHOR DECLARATIONS

**Conflict of Interest**

The authors have no conflicts to disclose

**Author Contributions**

**Po-Yen Chen**: Conceptualization; Data Curation; Formal Analysis; Investigation; Methodology; Software; Visualization; Writing/Original Draft Preparation

**Teruyasu Mizoguchi**: Supervision; Funding Acquistion; Writing/Review & Editing

DATA AVAILABILITY

The training data and the MLP model are openly available in Github, at https://github.com/nmdl-mizo/Finetuned-MACE-model.git, reference number 919981387.

ACKNOWLEDGMENT

This study was supported by the Ministry of Education, Culture, Sports, Science and Technology (MEXT) (Nos. 24H00042), and New Energy and Industrial Technology Development

Organization (NEDO). PYC would acknowledge the support of JST SPRING (Grant Number JPMJSP2108).

Figures

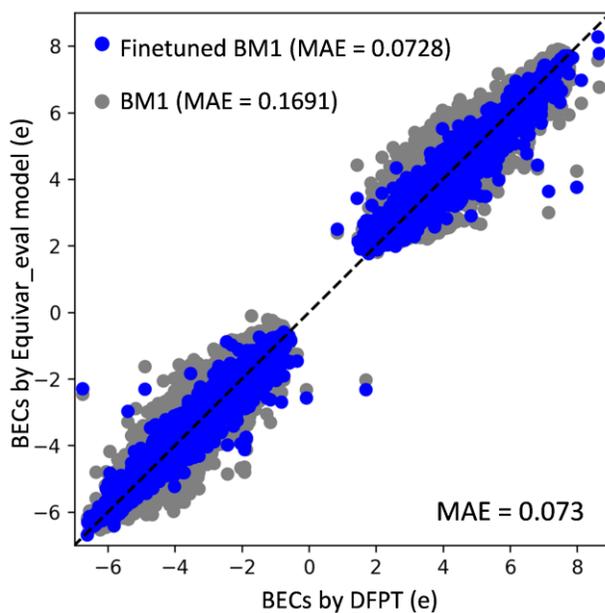

Figure 1. Comparison of the BECs obtained from the Equivar_eval model and DFPT calculations. The blue and gray dots represent the values predicted by the fine-tuned and pretrained BM1 models, respectively.

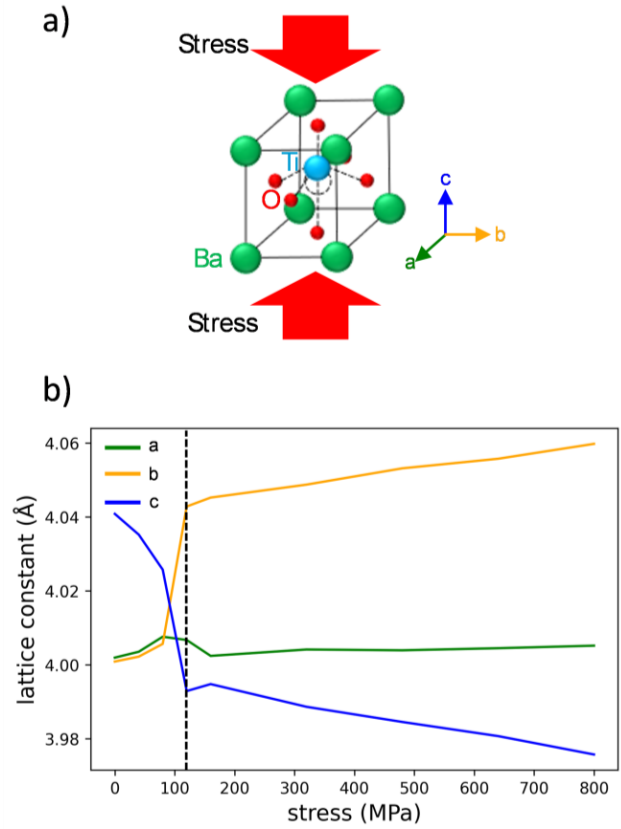

Figure 2 a) Schematic of the applied uniaxial compressive stress on the 8×8×8 BaTiO$_3$ supercell. b) Stress-dependent lattice constants at 250 K. The green, orange, and blue lines represent the a-, b-, and c-axis lattice constants, respectively, and the dashed line indicates the critical stress.

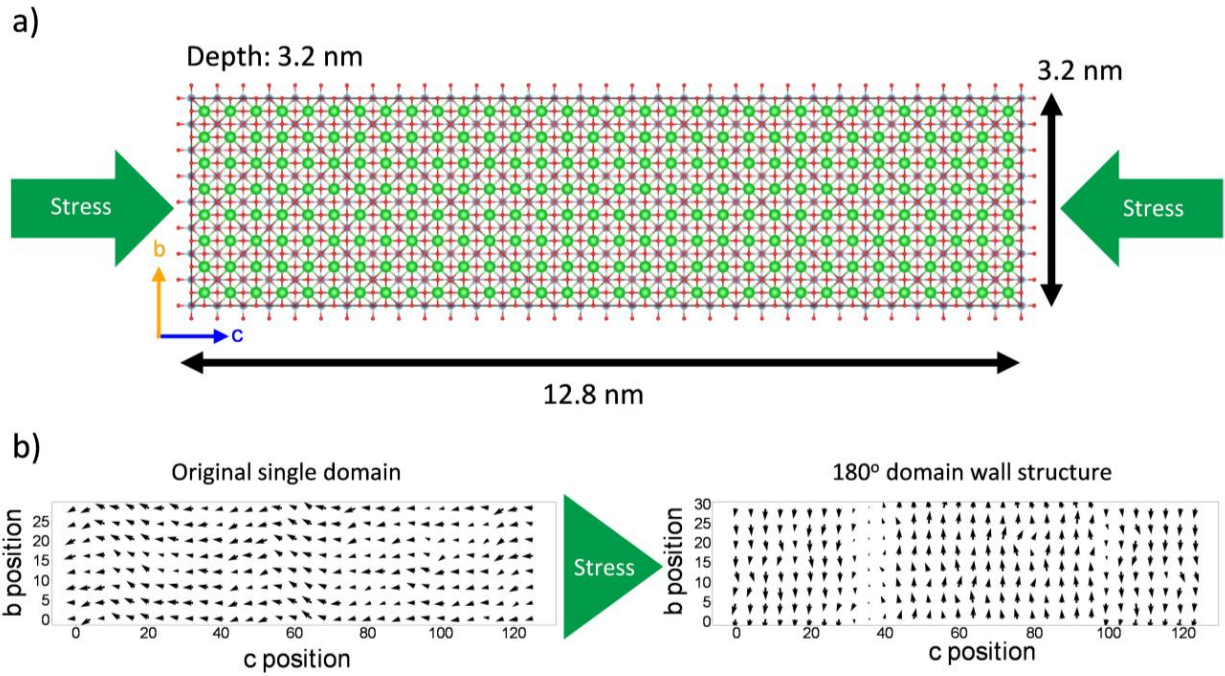

Figure 3 a) Schematic of the 32×8×8 supercell under applied stress. b) Ti-displacemnet change of the 32×8×8 supercell (10,240 atoms) under a uniaxial compressive stress of 800 MPa after 20 ps at 250 K. The arrows indicate the polarization orientations within the supercell.

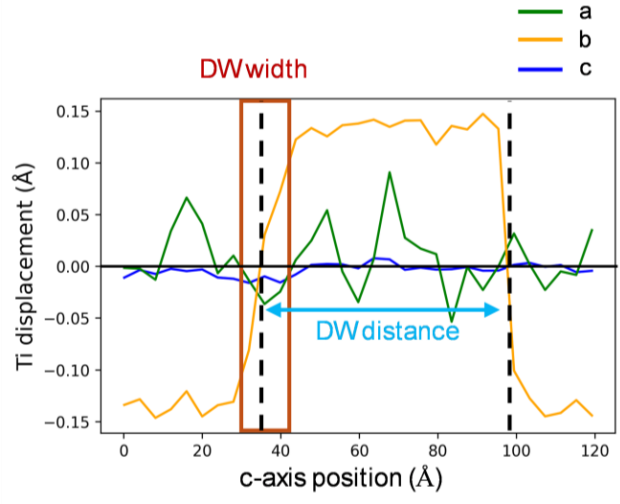

Figure 4 Position-dependent polarization along the a- (green), b- (orange), and c-axes (blue). The black dashed line indicates the DW position, the brown box highlights the region defining the DW width, and the cyan arrow shows the definition of the DW length.

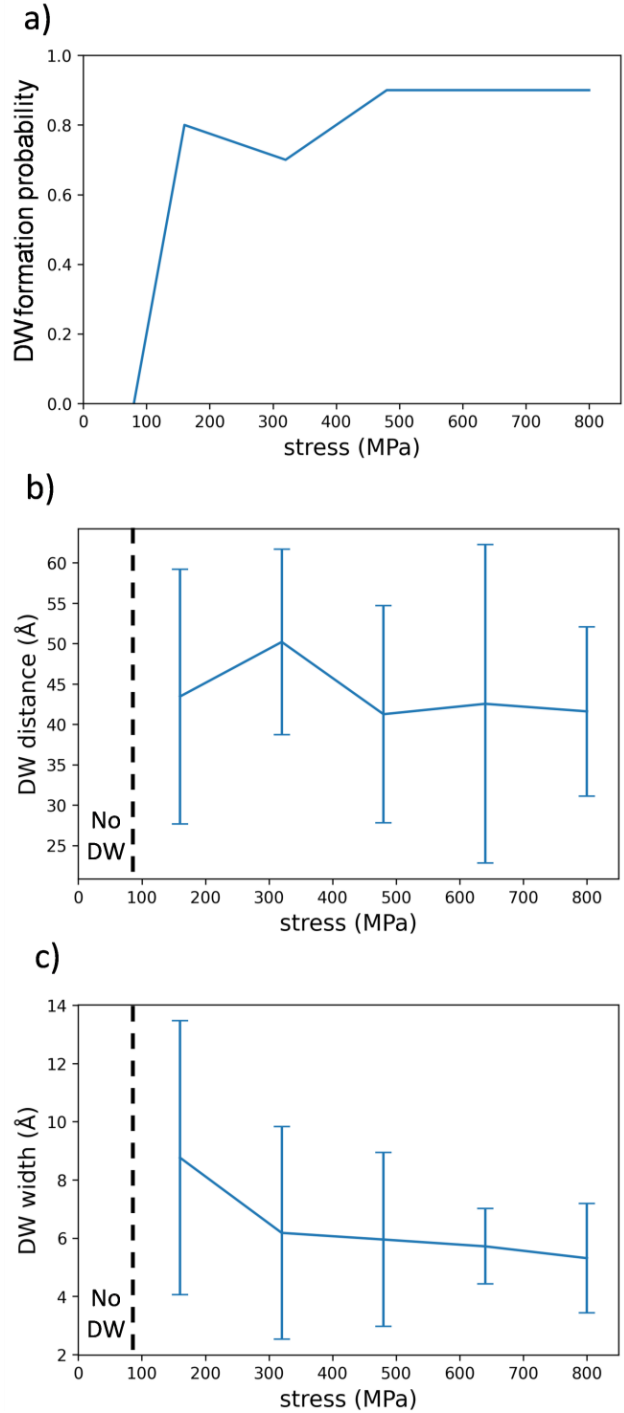

Figure 5 The stress dependence of (a) DW formation probability, (b) DW distance, and (c) DW width for 10 8×8×32 supercells at 250 K. Solid lines indicate the averaged values over the 10 simulations, and the error bars represent the standard deviation.

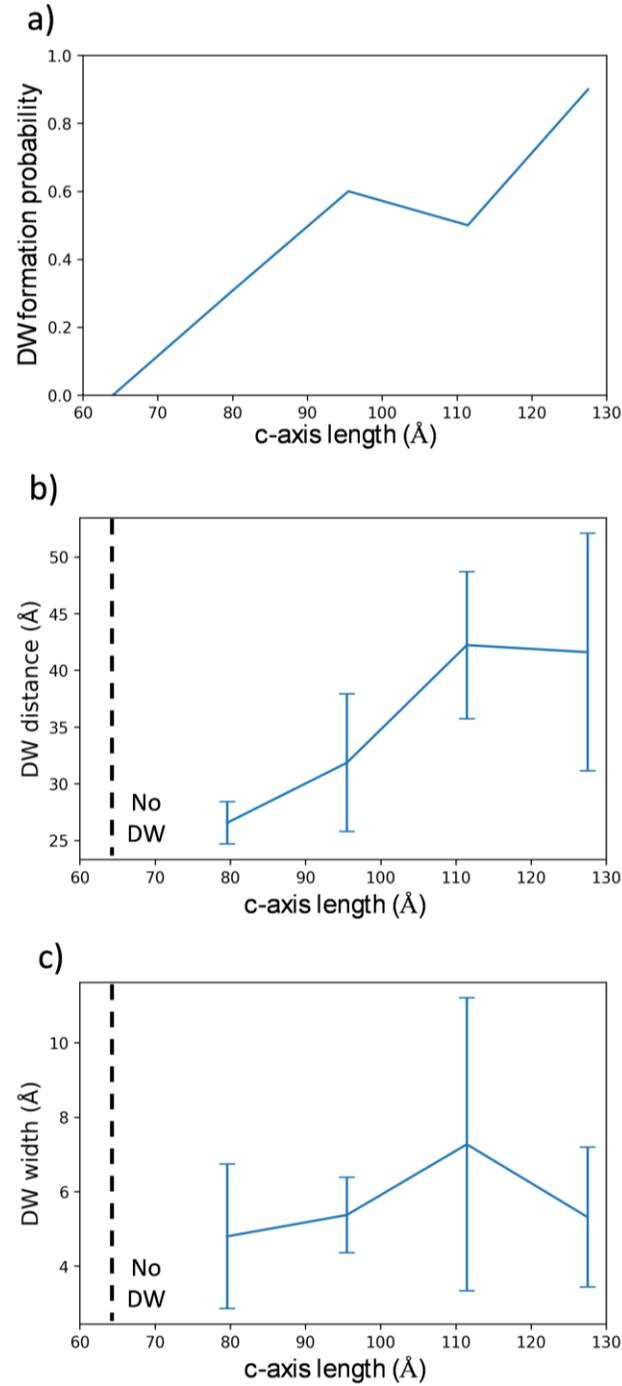

Figure 6 The c-axis length dependence of (a) DW formation probability, (b) DW distance, and (c) DW width for 64, 80, 96, 112, and 128 Å, corresponding to 16, 20, 24, 28, 32 supercells, along c-axis at 250 K under a uniaxial compressive stress of 800 MPa, with the ab-plane fixed as an 8×8

supercell. Solid lines indicate the averaged values over the 10 simulations for each supercell, and the error bars represent the standard deviation.

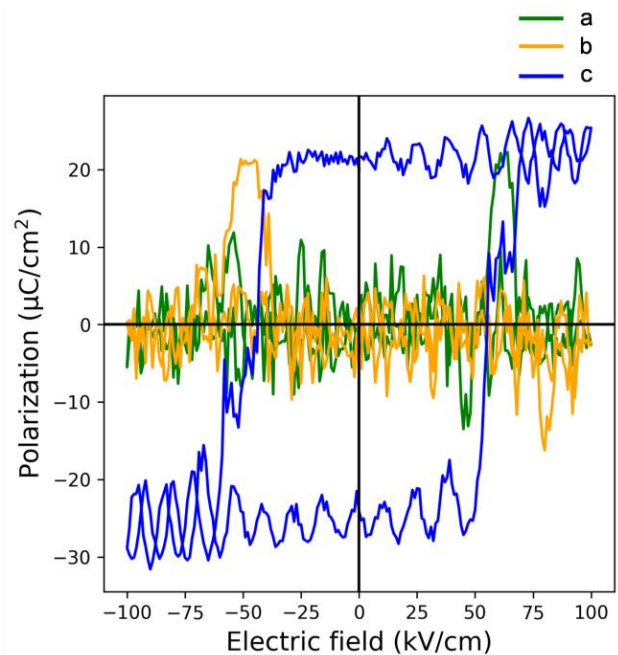

Figure 7 Hysteresis loops of the tetragonal 8×8×8 BaTiO$_3$ supercell at 250 K under zero mechanical stress. The green, orange, and blue lines represent the polarization along the a-, b-, and c-axes, respectively.

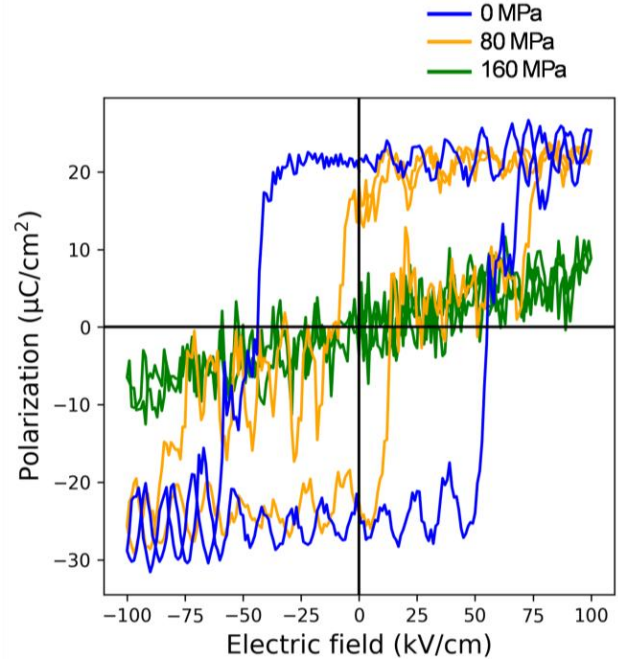

Figure 8 Hysteresis loops of the tetragonal 8×8×8 BaTiO$_3$ supercell at 250 K under different uniaxial compressive stresses. The green, orange, and blue lines represent the c-axis polarization under stresses of 0, 80, and 160 MPa, respectively.

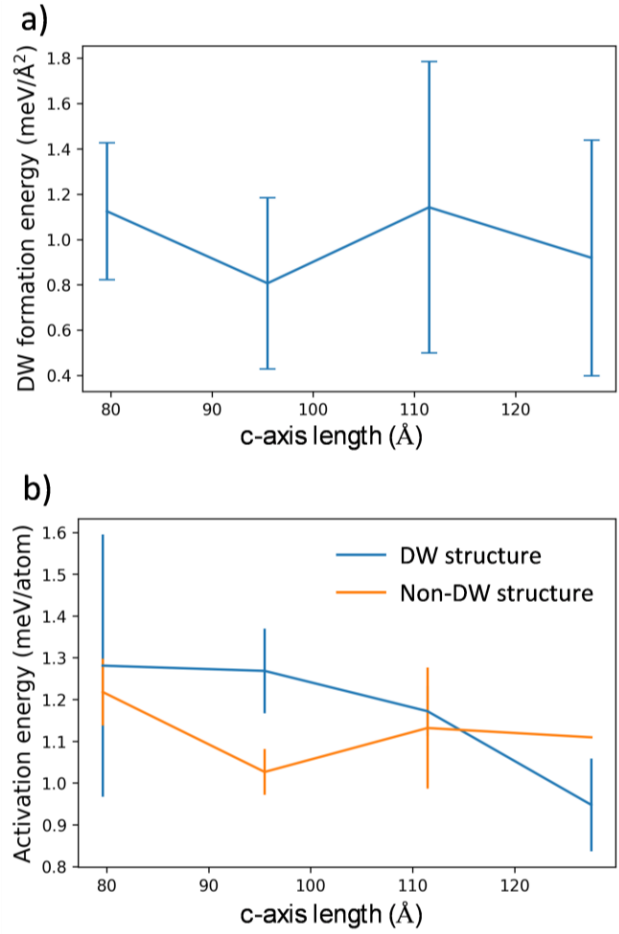

Figure 9 a) The c-axis length dependence of the DW formation energy, where the solid line represents the averaged values and the error bars denote the standard deviation. b) The c-axis length dependence of the activation energy for structures with (blue) and without (orange) DWs. The solid lines represent the averaged values, and the error bars indicate the standard deviation.

# Supporting information

Effect of uniaxial compressive stress on polarization switching and domain wall formation in tetragonal phase BaTiO$_3$ via machine learning potential

*Po-Yen Chen[1], Teruyasu Mizoguchi[1,2].*

AUTHOR ADDRESS

[1]Department of Materials Engineering, the University of Tokyo, Tokyo, Japan

[2]Institute of Industrial Science, the University of Tokyo, Tokyo, Japan.

AUTHOR INFORMATION

**Corresponding Author**

poyen@iis.u-tokyo.ac.jp, teru@iis.u-tokyo.ac.jp

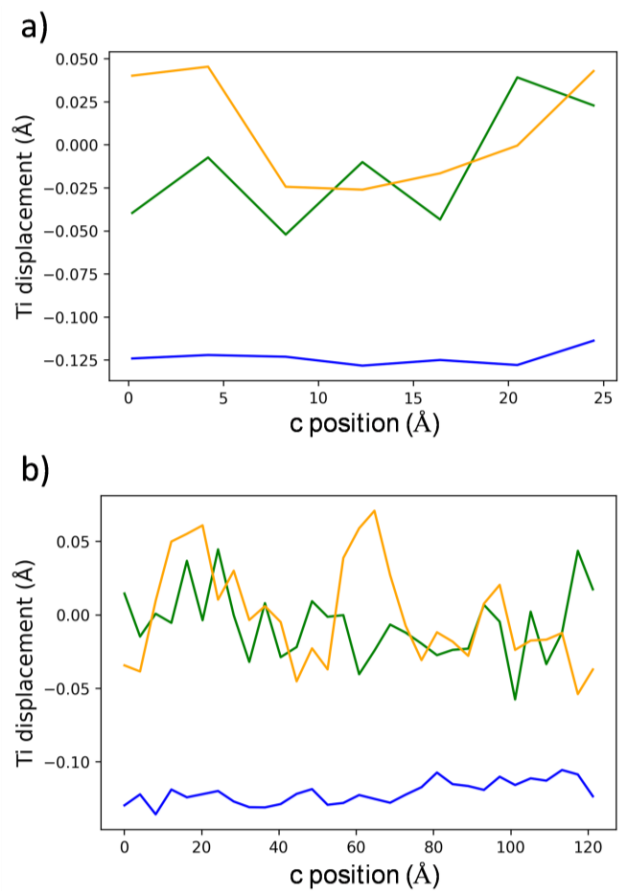

**Figure S1** The c-position dependent Ti-displacement for a) 8x8x8 supercell and b) 32x8x8 supercell at 250K.

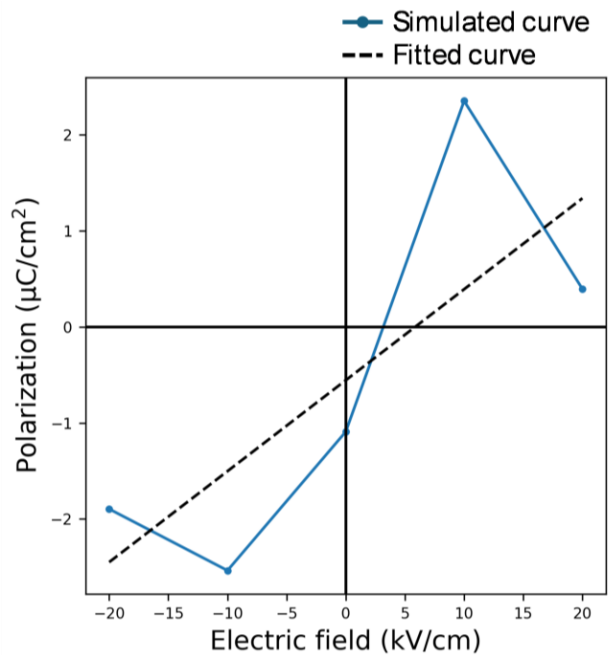

**Figure S2** The P-E curve for the linear response region for the BaTiO$_3$ along a-axis. The blue curve is the data from the MD simulation, and the dashed line is the fitted curve.

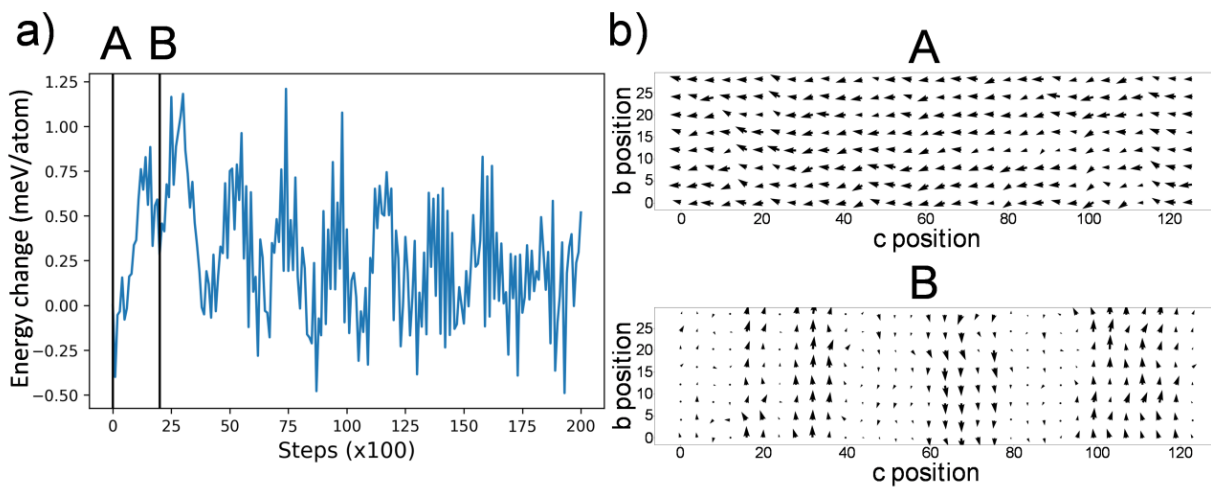

**Figure S3** a) The step-dependent energy change during the polarization switching process. b) The polarization of the snapshot for the initial structure (A) and the structure at the 2000$^{th}$ MD step (B).